\documentclass{jpsj2}
\title
{Thermodynamic and Transport Properties of CeMg$_{2}$Cu$_{9}$ under Pressure}
\author
{Masakazu {\sc Ito} \footnote{E-mail: showa@hiroshima-u.ac.jp}, 
 Koji {\sc Asada},
 Yuko {\sc Nakamori}$^{1}$ \footnote{Present address: Institute for Materials Research, Tohoku University, Sendai 980-8577, Japan},
 Jyun'ya {\sc Hori},
 Hironobu {\sc Fujii}$^{2}$,
 Fumihiko {\sc Nakamura},
 Toshizo {\sc Fujita}
 and
 Takashi {\sc Suzuki} \footnote{E-mail: tsuzuki@hiroshima-u.ac.jp}.
}
\inst{
  Department of Quantum Matter, ADSM, 
Hiroshima University, Higashi-Hiroshima 739-8530 Japan\\
$^1$ Faculty of Integrated Arts $\&$ Sciences, Hiroshima University, Higashi-Hiroshima 739-8526 Japan\\
$^2$ Natural Science Center for Basic Research and Development,  
 Materials Science Center, Hiroshima University, Higashi-Hiroshima 739-8526 Japan\\
}
\recdate
{\today}
\abst
{
We report the transport and thermodynamic properties under hydrostatic pressure in the antiferromagnetic Kondo compound CeMg$_{2}$Cu$_{9}$ with a two-dimensional arrangement of Ce atoms.
Magnetic specific heat $C_{\rm mag}$($T$) shows a Schottky-type anomaly around 30 K originating from the crystal electric field (CEF) splitting of the 4f state with the first excited level at $\Delta_{1}/k_{\rm B} =$ 58 K and the second excited level at $\Delta_{2}/k_{\rm B} =$ 136 K  from the ground state.
 Electric resistivity shows a two-peaks structure due to the Kondo effect on each CEF level around $T_{1}^{\rm max}$ = 3 K and $T_{2}^{\rm max}$ = 40 K. These peaks merge around 1.9 GPa with compression.   
With increasing pressure, N$\acute{\rm e}$el temperature $T_{\rm N}$ initially increases and then change to decrease. $T_{\rm N}$ finally disappears at the quantum critical point $P_{\rm c}$ = 2.4 GPa.  }
\kword
{CeMg$_{2}$Cu$_{9}$, Specific heat, Electric resistivity, Kondo effect, Pressure effect, Crystal field effect
}
\begin{document}
\sloppy
\maketitle
\section{Introduction}
In recent years, the studies on pressure effects on Ce based-inter metallic compounds have been carried out intensively, since the number of interesting physical properties, such as valence fluctuation,\cite{Nakajima_Sm4Bi3} non-fermi liquid\cite{Lohneysen, Umeo} and superconductivity,\cite{Grosche_CePd2Si2, Movshovich_CeRh2Si2, Vargoz_CeCu2, Walker_CeIn3} emerge by mean of the control physical parameters for the correlated electron systems. 
Especially, the discovery of superconductivity in CeRhIn$_{5}$,\cite{Hegger_CeRhIn5} which has two dimensional alignment of the Ce atoms, attracts much attention because CeRhIn$_{5}$ has potential to provide a unique opportunity to investigate the relation between superconductivity and not only magnetism but also structural dimensionality.
We believe that the two-dimensional crystal structure plays an important role for the appearance of unusual properties in the Ce-based inter metallic compounds. 
CeMg$_{2}$Cu$_{9}$ is one of the good candidate materials for investigating the role of the two-dimensionality with the strong correlation. 
This compound was firstly synthesized by Y. Nakamori $et\ al$. \cite{Nakamori}who reported that the crystal structure is the hexagonal CeNi${_3}-$type ( space group $P{\rm6_{3}}/mmc$ ).
The structure of this unitcell is built up by stacking of alternating MgCu$_{\rm2}$ Laves-type and rare earth based CeCu$_{\rm5}$-type layers along $c$-axis.
The distance between the nearest Ce atoms along the $c$-axis is $\sim$ 8.6 $\AA$  which is 1.7 times larger than that in the $c$ plane ($\sim$ 5.1 $\AA$).
One can expect a two-dimensional electronic properties reflecting the two-dimensional crystal structure. 
In this paper, we report specific heat and electrical resistivity on CeMg$_2$Cu$_9$ under hydrostatic pressures.
If in CeRhIn$_{5}$, the appearance of superconductivity by compression is related to the two dimensionality, it is interesting to search superconductivity in CeMg$_{2}$Cu$_{9}$ by compression.
\par
\section{Experimental}
 Polycrystalline samples of CeMg$_{2}$Cu$_{9}$ and LaMg$_{2}$Cu$_{9}$ were prepared by melting stoichiometric amount of consistent metals at 1200$^\circ$C in the 0.5 MPa Ar atmosphere in a Mo-crucible.
  Specific heat $C_P$ measurements were carried out by a conventional adiabatic heat-pulse method.
 A piston-cylinder Cu-Be clamp cell, which contains Apiezon-J oil as pressure-transmitting oil, was adopted to measure $C_P$($T$) under pressure $P$ up to 1.0 GPa. Temperature was decreased down to 0.5 K using a $^{\rm 3}$He refrigerator.
  Electrical resistivity $\rho $ were measured by a standard four-probe method in the temperature range between 2  and 300 K. 
  We utilized two types of pressurization technique according to the range of pressure. A cubic anvil device was used for the $P$ range between 3.0 and 8.0 GPa.\cite{Mori,Hori} 
 For the $\rho $($T$) measurement in the range between 0.1 MPa and 2.3 GPa, we used a piston-cylinder type tungsten-carbide (WC) clamp cell with the equal volume mixture of Fluorinert FC70 and FC77 as the pressure transmitter. 
Dilution refrigerator was used to measure $\rho $($T$) from low temperature ( $\sim$ 0.1 K), for 2.3 GPa.
\section{Results}
\subsection{Specific heat}
  Figure 1(a) shows the temperature $T$ dependence of specific heat $C_P$($T$) of CeMg$_{2}$Cu$_{9}$ at 10$^{-4}$ Pa from 0.5 to 50 K together with that of the isostructural nonmagnetic LaMg$_{2}$Cu$_{9}$.  
  A large peak which indicates the antiferromagnetic transition is observed at $T_{\rm N} $= 2.5 K. 
  Magnetic specific heat $C_{\rm mag}$($T$) is obtained by subtraction of $C_P$($T$) of LaMg$_{2}$Cu$_{9}$ from that of CeMg$_{2}$Cu$_{9}$, which is shown in Fig. 1 (b).   
We roughly estimated that the Sommerfeld coefficient is not less than 117 mJ/K$^{2}$mol from the value of $C_{\rm mag}$/$T$ at $T$ = 5 K. 
  A Schottky-type anomaly is found around 30 K, indicating the crystal electric field ( CEF ) splitting of the 4$f$-electronic levels at the Ce ion site.
  In the hexagonal symmetry, the sixfold degenerate state of 4$f$ for Ce$^{3+}$ splits into the three Kramers doublets. The Schottky anomaly is theoretically given by
\begin{eqnarray}
C_{\rm Sc}(T) = \frac{R}{T^{2}}\left[  \right. \frac{\Delta_{1}^{2}  e^{-\Delta_{1}/k_{\rm B}T}+\Delta_{2}^{2}  e^{-\Delta_{2}/k_{\rm B}T}}{k_{\rm B}^2Z(T)}\nonumber\\
-\left( \frac{\Delta_{1}  e^{-\Delta_{1}/k_{\rm B}T}+\Delta_{2}  e^{-\Delta_{2}/k_{\rm B}T}}{k_{\rm B}Z(T)} \right)^{2} \left.  \right], 
\end{eqnarray}
with $Z(T)=1+\exp (-\Delta_{1}/k_{\rm B}T)+\exp (-\Delta_{2}/k_{\rm B}T)$, where $R$, $k_{\rm B}$, $\Delta_{1}$ and $\Delta_{2}$ are the gas constant, Bolzman factor, the first and second excited CEF energies measured from the ground level, respectively. 
The experimental result can be reproduced by eq. (1) with the parameters $\Delta_{1}/k_{\rm B}$ = 58 K and $\Delta_{2}/k_{\rm B}$ = 136 K as shown by the broken curve in Fig. 1(b). 
\begin{figure}
\begin{center}
\includegraphics[height=150mm,clip]{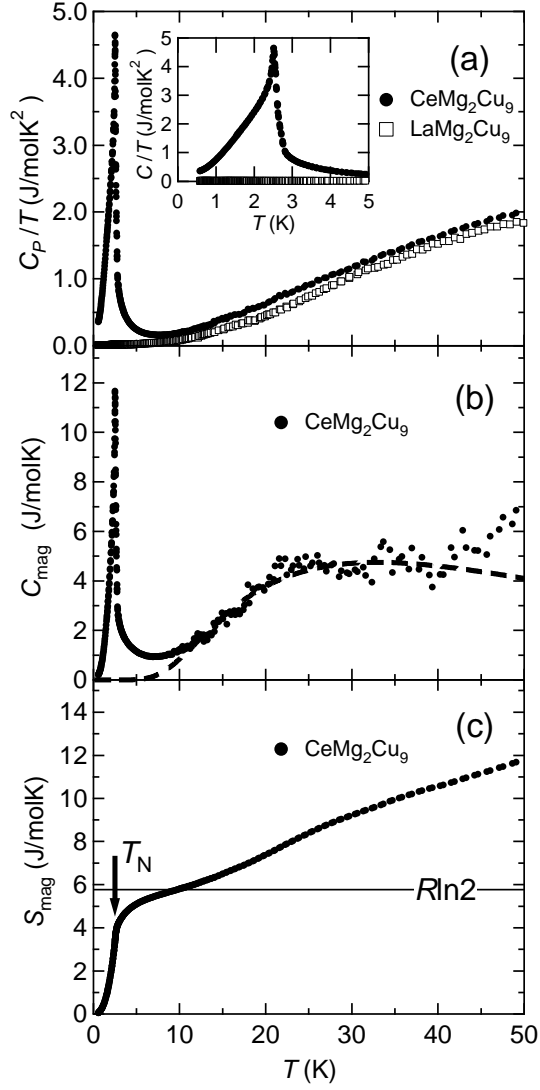}
\caption{(a)Temperature dependence of specific heat divided by temperature $C_{\rm P}/T$ of CeMg$_{2}$Cu$_{9}$ shown by the closed circles and LaMg$_{2}$Cu$_{9}$ in 0.5 $\le  T \le $ 50 K at $P$ = 10$^{-4}$ Pa shown by the open rectangles. (b)Temperature dependence of magnetic specific heat $C_{\rm mag}$. The fitting result with eq. (1) is drawn by a broken curve. (c)Temperature dependence of the magnetic entropy $S_{\rm mag}$.}
\label{fig:1}
\end{center}
\end{figure}
The magnetic entropy $S_{\rm mag}$($T$) is obtained by the relation
\begin{eqnarray}
S_{\rm mag}(T) = \int_{0}^{T} \frac{C_{\rm mag}(T)}{T}dT, 
\end{eqnarray}
and shown in Fig. 1 (c). 
The released $S_{\rm mag}$ at $T_{\rm N}$ is about 60$\%$ of $R$ln2, and remaining entropy ( 0.4$R$ln2 ) is recovered by 10 K. 
This suggests that the Kondo-compensated ordered moments are formed in the low temperature range. 
The Kondo interaction can be responsible for the entropy transfer to higher temperature.\cite{Bauer_CeCu5, Bredl}
This is reflected in tail of $C_{\rm mag}$ above $T_{\rm N}$.
   Figure 2 (a) displays temperature dependence of $C_P$/$T$ at various pressures up to 0.91 GPa from 0.5 to 4 K.
Anomalies at $T_{\rm N}$ have nearly equivalent sharpness and amplitude even for 0.91 GPa.     
The value of released $S_{\rm mag}$ at $T_{\rm N}$ is not change by pressure below 0.91 GPa as shown in Fig. 2 (b).
This suggests that the Kondo-compensated ordered moments are still formed in 0.91 GPa. 
Figure 3 shows pressure dependence of $T_{\rm N}$. With increasing $P$, $T_{\rm N}$ slightly increases up to 2.58 K at 0.89 GPa, and changes to decrease. 
   \begin{figure}
\begin{center}
\includegraphics[height=140mm,clip]{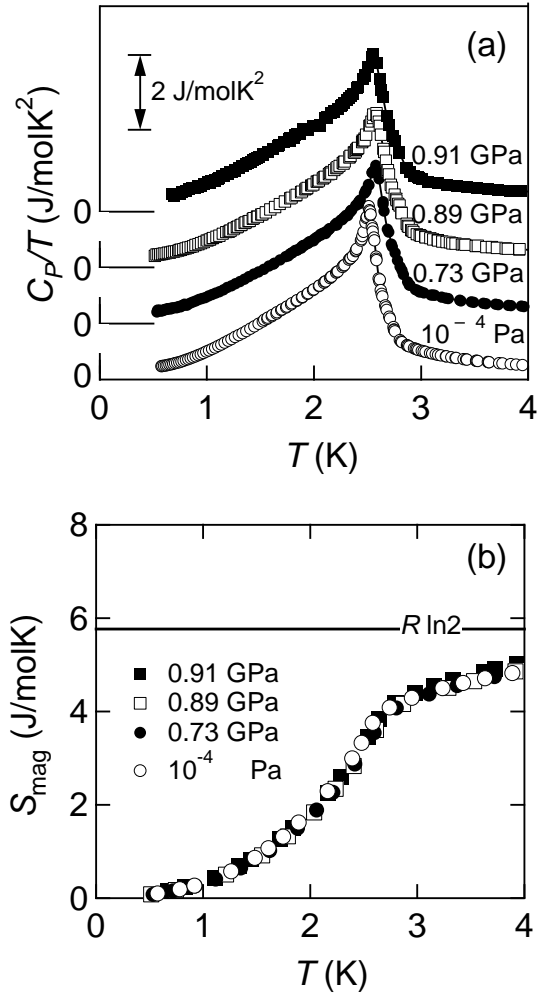}
\caption{Temperature dependence of (a) $C_P$/$T$ and (b) $S_{\rm mag}$ at various pressures up to 0.91 GPa in $T \le $4.0 K.}
\label{fig:2}
\end{center}
\end{figure}
   \begin{figure}
\begin{center}
\includegraphics[height=65mm,clip]{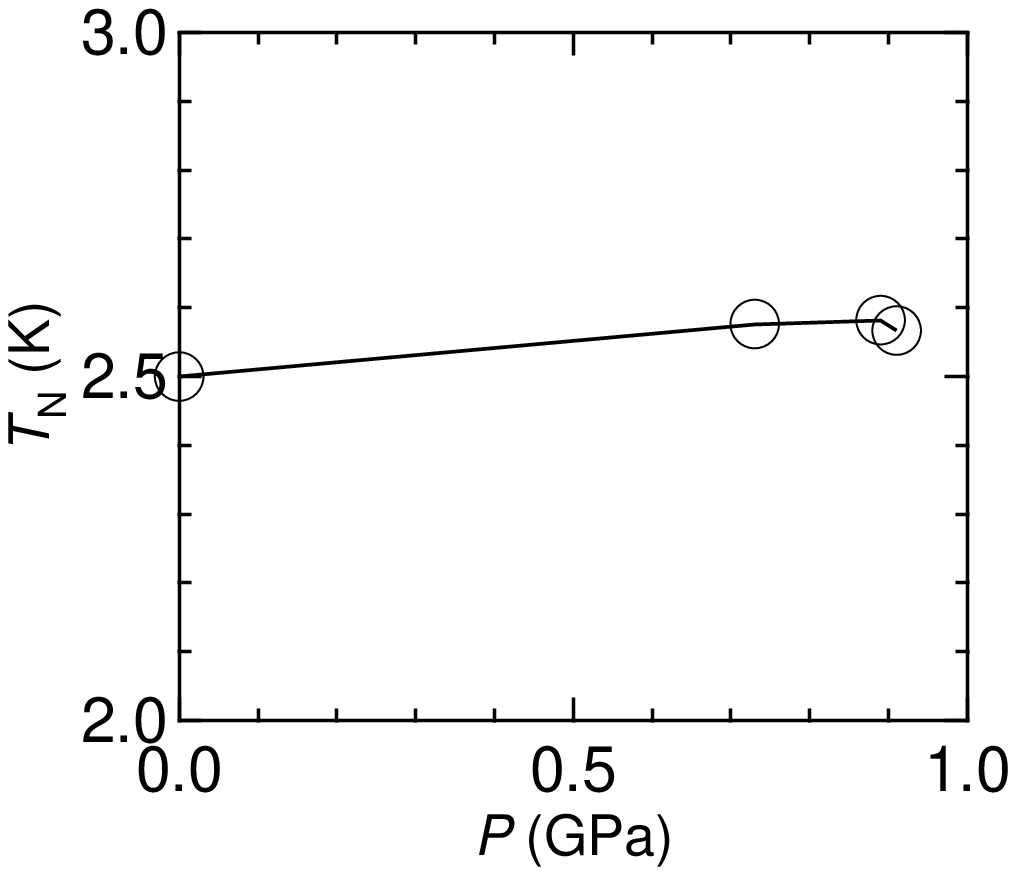}
\caption{Pressure dependence of $T_{\rm N}$ obtained from $C_{\rm mag}(T)$. The solid line is guide to the eyes. }
\label{fig:3}
\end{center}
\end{figure}
\subsection{Electric resistivity}
 Temperature dependencies of the electrical resistivity $\rho$($T$) of CeMg$_{2}$Cu$_{9}$ and LaMg$_{2}$Cu$_{9}$ are plotted in Fig. 4 (a). 
The drop at 2.5 K, which was defined as the maximum point of $\partial \rho_{\rm mag}/ \partial T$, is due to the antiferromagnetic ordering.
The magnetic contribution $\rho_{\rm mag}$($T$) to the resistivity of CeMg$_{2}$Cu$_{9}$ was obtained by the subtracting  $\rho_{\rm{LaMg_{2}Cu_{9}}}$ from $\rho_{\rm{CeMg_{2}Cu_{9}}}$ and is shown in the semi-logarithmic scale in Fig. 4(b).
$\rho_{\rm mag}$($T$) shows the double-peaked structure around $T_{1}^{\rm max}$ = 3 K and $T_{2}^{\rm max}$ = 40 K. Similar feature were reported for some of Ce-based heavy fermion (HF) compounds, for example, CeAl$_{2}$,\cite{Onuki_CeAl2} CePdSi$_{2}$\cite{Lu_CePdSi2} and CePb$_{3}$\cite{Suzuki_CePb3}. 
By analogy of the cases for these HF compounds, the resistivity peak at $T_{1}^{\rm max}$ may originate from the Kondo scattering with the energy scale of Kondo temperature $T_{\rm K}$ for the ground state and $T_{2}^{\rm max}$ from the Kondo scattering with higher-Kondo temperature $T_{\rm K}^{h}$ on the whole CEF state.
For the three Kramers doublets system with the energy gaps $\Delta$$_1$ ( $>$ $T_{\rm K}$ ) and $\Delta$$_2$, the relation among $T_{\rm K}$, $T_{\rm K}^{h}$, $\Delta$$_1$ and $\Delta$$_2$ is described as\cite{Hanzawa}
\begin{eqnarray}
(k_{\rm B}T_{\rm K}^{h})^{3}=\Delta_{1}\Delta_{2}(k_{\rm B}T_{\rm K}).
\end{eqnarray}
From the above equation we calculated $T_{\rm K}$ $\sim$ 8 K with $\Delta_{1}/k_{\rm B}$ and $\Delta_{2}/k_{\rm B}$ determined from specific heat, where we assumed $T_{\rm K}^{h}$ = $T_{2}^{\rm max}$. 
Logarithmic temperature dependencies, which are indicated by the broken lines $m_{1}$ and $m_{2}$ in Fig. 4(b), can be understood by the theoretical model by Cornut and Coqblin.\cite{Cornut} 
According to their model, $-$ln($T$) dependencies arise from the  Kondo effect in the CEF levels, and $\rho_{\rm mag}$($T$) is given by 
\begin{eqnarray}
\rho_{\rm mag} = \rho_{sd} + 2AJ_{cf}^{3}n_{f} \frac{\lambda ^{2} - 1}{2J + 1} \ln (k_{\rm B}T / D), 
\end{eqnarray}
where $\rho_{sd}$ is the spin disorder resistivity, 
$A$ is the constants, $n_{f}$ is the density of states at the Fermi level, $J_{cf}$ ( $<$ 0 ) is the exchange interaction between the conduction and localized 4$f$ electrons, $J$ is total angular momentum of 5/2 for Ce$^{3+}$, and $D$ is the effective band width. 
The parameter $\lambda$ is the number of thermally accessible states, for example, $\lambda$ = 2 for the ground state doublet as the low $T$ limit and 6 (= 2$J$ + 1) as the high $T$ limit, respectively. \cite{Cornut,Francillon_CeAl2} 
The ratio of the logarithmic slopes $m_{2}/m_{1}$ is 4.5, and this value is close to $\rho_{\rm mag}(\lambda = 4)/\rho_{\rm mag}(\lambda = 2)$ (= 15/3) obtained from eq. (1) by assuming that the contribution of $\rho_{sd}$ is small.
This suggests that $-$ln($T$) dependence $m_{1}$ and $m_{2}$ arises from the Kondo effect at the ground state $(\lambda = 2)$
and the first excited state $(\lambda = 4)$, respectively. 
\cite{Cornut}
\begin{figure} 
\begin{center}
\includegraphics[height=65mm,clip]{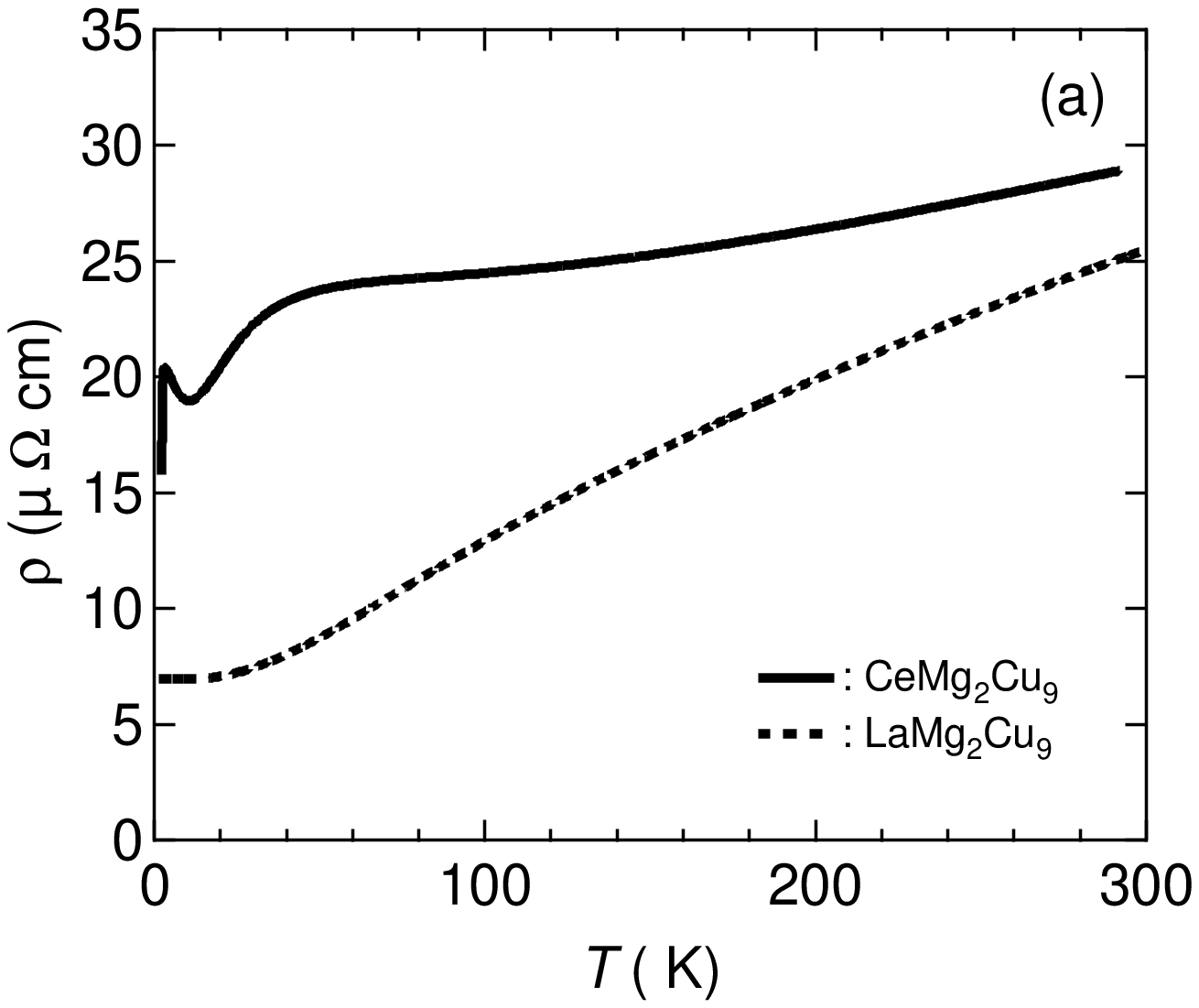}
\includegraphics[height=65mm,clip]{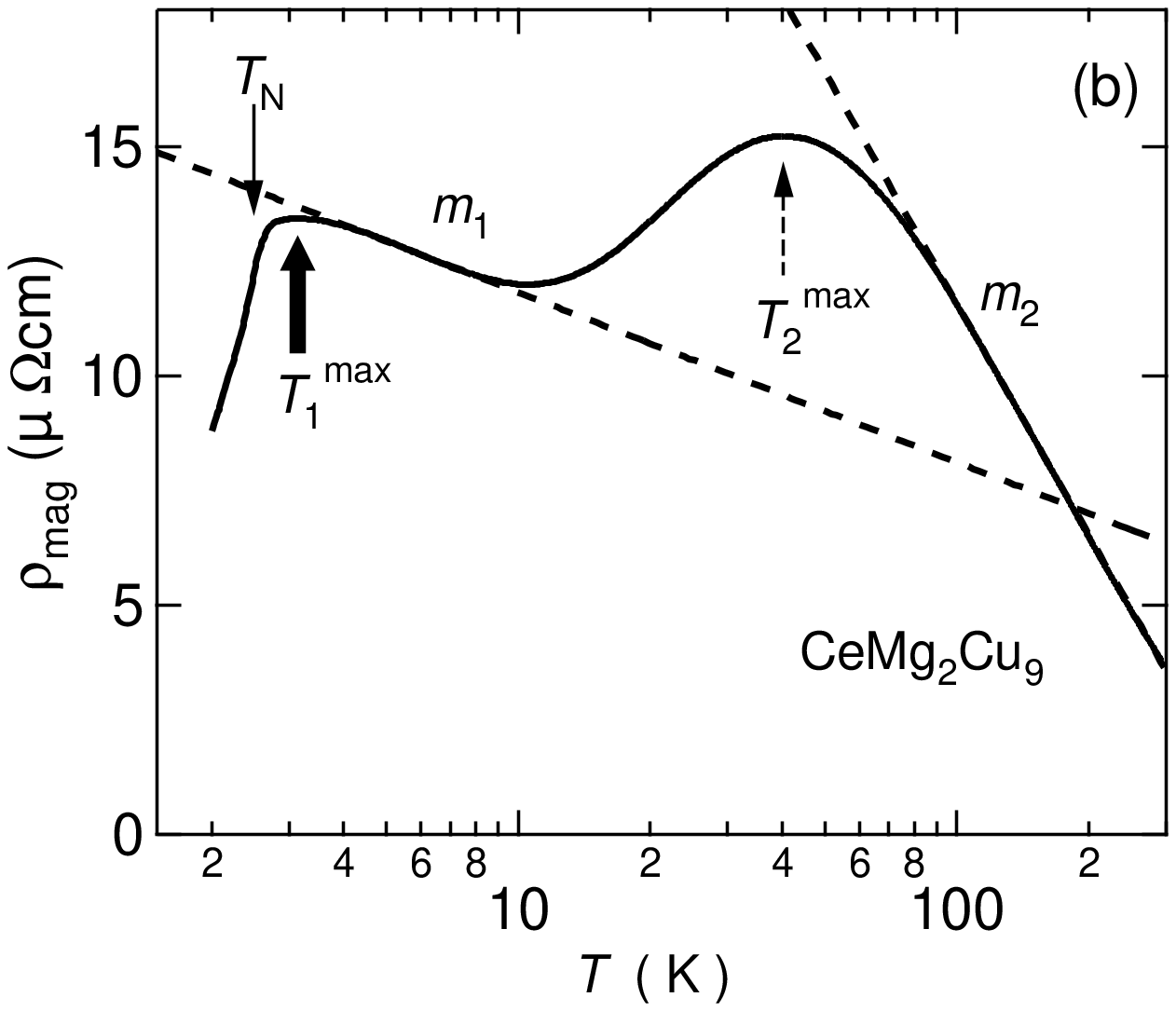}
\caption{(a)Temperature dependence of electrical resistivity $\rho$($T$) of CeMg$_{2}$Cu$_{9}$ presented by a solid curve and LaMg$_{2}$Cu$_{9}$ presented by a broken curve between 1.5 and 300 K at $P$ = 10$^{-4}$ Pa. (b) Semi-logarithmic plot of magnetic resistivity $\rho_{\rm mag}$($T$).
The broken lines are guides to the eyes, indicating $-$ln$T$ dependence with slope $m_{\rm 1}$ and $m_{\rm 2}$ of $\rho_{\rm mag}$($T$).
We defined N$\acute{\rm e}$el temperature  $T_{\rm N}$  as the maximum point of $\partial \rho_{\rm mag}/ \partial T$.
}
\label{fig:4}
\end{center}
\end{figure}
  $\rho_{\rm mag}$($T$) measured by using the WC clamp cell in the pressure range 0.1 MPa$ \le  P \le $ 2.3 GPa are shown in Fig. 5 (a) for some representative pressures.  
As expected from $C_P$($T$) above 0.89 GPa,  $T_{\rm N}$ decreases with compression. $T_{\rm N}$ reduced to 0.22 K at 2.3 GPa as shown in the inset of Fig. 5 (a).
$T_{1}^{\rm max}$ is almost independent of $P$, whereas $T_{2}^{\rm max}$ moves to lower temperatures with the ratio $\partial T_{2}^{\rm max}/\partial P \sim - 17$ K/GPa and the two peaks merge at $T^{\rm max}$ around 1.9 GPa. 
$T^{\rm max}$ moves to higher temperature with $\partial T^{\rm max}/\partial P \sim 20$ K/GPa above 3 GPa by compression as shown in Fig. 5 (b) which is plotted $\rho_{\rm mag}$($T$) under $P$ generated by the cubic anvil device in 3.0 $ \le  P \le $ 8.0 GPa. 
The negative $\partial T_{2}^{\rm max}/\partial P$ observed below 1.6 GPa is unusual. 
According to the pressure studies on the Ce-based HF compounds,  $\partial|J_{cf}n_{f}|$/$\partial P$ is to be positive,\cite{ Kagayama} 
because  $T_{\rm K}$ and $T_{\rm K}^{h}$ are in proportion to $\exp (-1/|J_{cf}n_{f}|)$,  $T_{1}^{\rm max}$ and $T_{2}^{\rm max}$ should increase with compression as observed in CePb$_3$.\cite{Suzuki_CePb3} 
It is not clear why the $\partial T_{2}^{\rm max}/\partial P$ is  the negative value in CeMg$_{2}$Cu$_{9}$. 
This unusual behavior is seen in pressure dependence of $\rho_{\rm mag}$($T$) of CeRhIn$_{5}$ which is a pressure-induced superconductor with the 2D structure. 
\begin{figure}
\begin{center}
\includegraphics[height=180mm,clip]{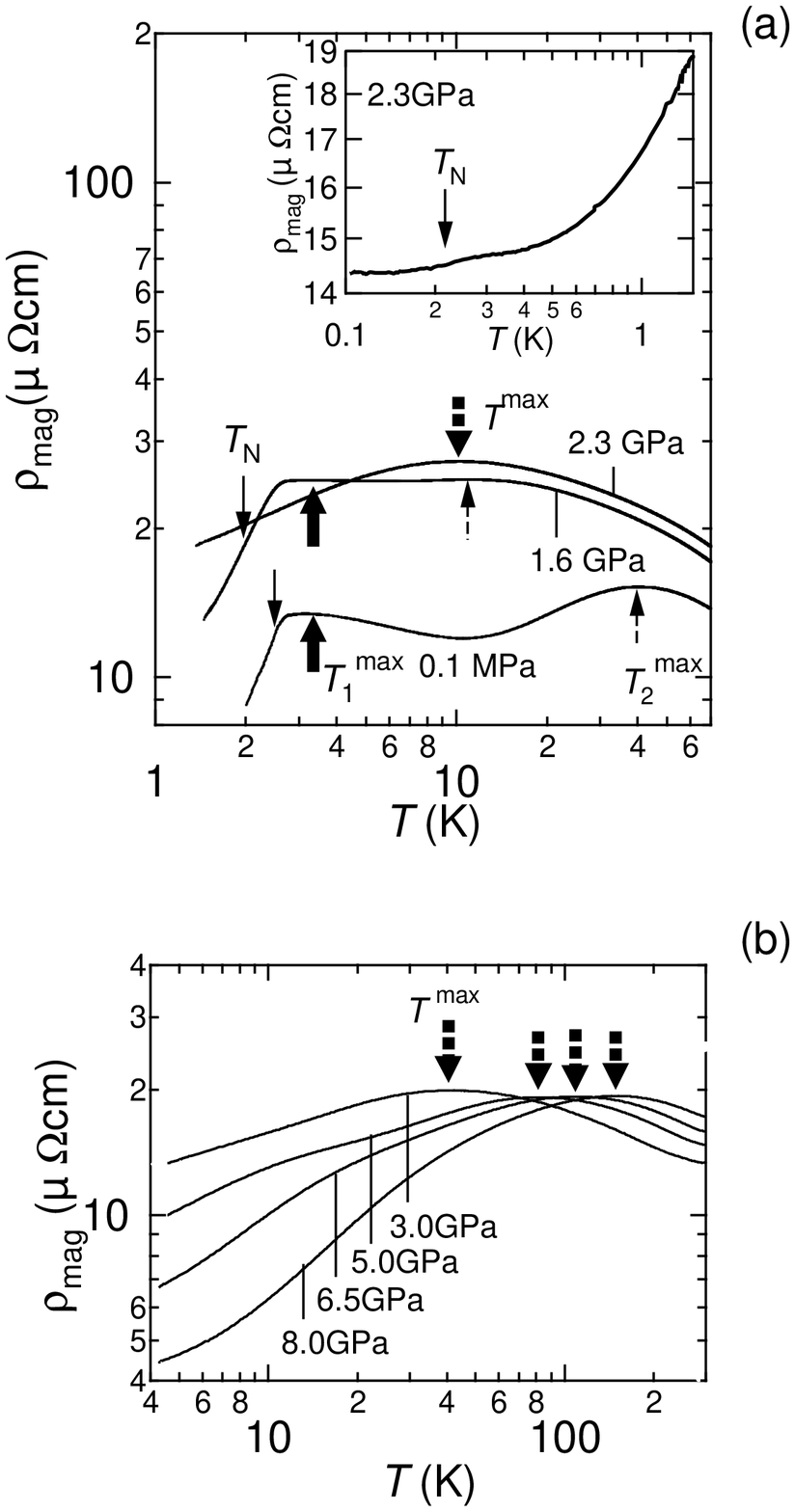}
\caption{Temperature dependence of electrical resistivity $\rho_{\rm mag}$($T$) under pressures. (a) $\rho_{\rm mag}$($T$) in the range of 1.5 $\le  T \le $ 70 K and 0.1 MPa $\le  P \le  $ 2.3 GPa measured using the WC clamp cell.  Inset shows $\rho_{\rm mag}$($T$) at 2.3 GPa in the range of 0.1 $\le  T \le $ 2 K measured using the WC clamp cell in a dilution refrigerator. (b) $\rho_{\rm mag}$($T$) in the range of 4.2 $\le  T \le $ 300 K and 1. 5 $\le  P \le  $ 8.0 GPa measured using the cubic-anvil device.  The thin, thick, thin-broken and thick-broken arrows show the positions of N$\acute{\rm e}$el temperature $T_{\rm N}$,  characteristic temperatures $T_{1}^{\rm max}$, $T_{2}^{\rm max}$ and $T^{\rm max}$, respectively. }
\label{fig:5}
\end{center}
\end{figure}
 
The effects of compression on $T_{\rm N}$, $T_{1}^{\rm max}$, $T_{2}^{\rm max}$ and $T^{\rm max}$ are summarized in Fig. 6.  
We can easily estimate that $T_{\rm N}$ disappears at the critical pressure $P_{c}$ = 2.4 GPa, which is so called quantum critical point (QCP). 
This value is very close to the value (2.5 GPa) obtained from the resistivity measurement under pressure, which reported by Nakawaki $et\ al$.\cite{Nakawaki_CeMg2Cu9}. 
A large number of studies on QCP in the Ce-based compounds have been carried out, and some of them, for example on CePd$_{2}$Si$_{2}$,\cite{Grosche_CePd2Si2} CeRh$_{2}$Si$_{2}$\cite{Movshovich_CeRh2Si2} and CeIn$_{3}$\cite{Walker_CeIn3}, find superconductivity at vicinity of the $P_{c}$ in the low temperature range .
 In the case of CeMg$_{2}$Cu$_{9}$, we have not observed pressure-induced superconductivity around $P_{c}$ so far. 
Many Ce-based HF superconductors have shown that pairing symmetry is anisotropic.\cite{Ishida_CeRh2Si2, Fisher_CeRhIn5, Izawa_CeRhIn5}
When the system has an anisotropic pairing symmetry, superconducting transition temperature $T_{\rm c}$ is strongly suppressed by a small amount of impurity.
We continue the search of the pressure induced superconductivity with high-quality CeMg$_{2}$Cu$_{9}$.
\begin{figure}
\begin{center}
\includegraphics[height=70mm,clip]{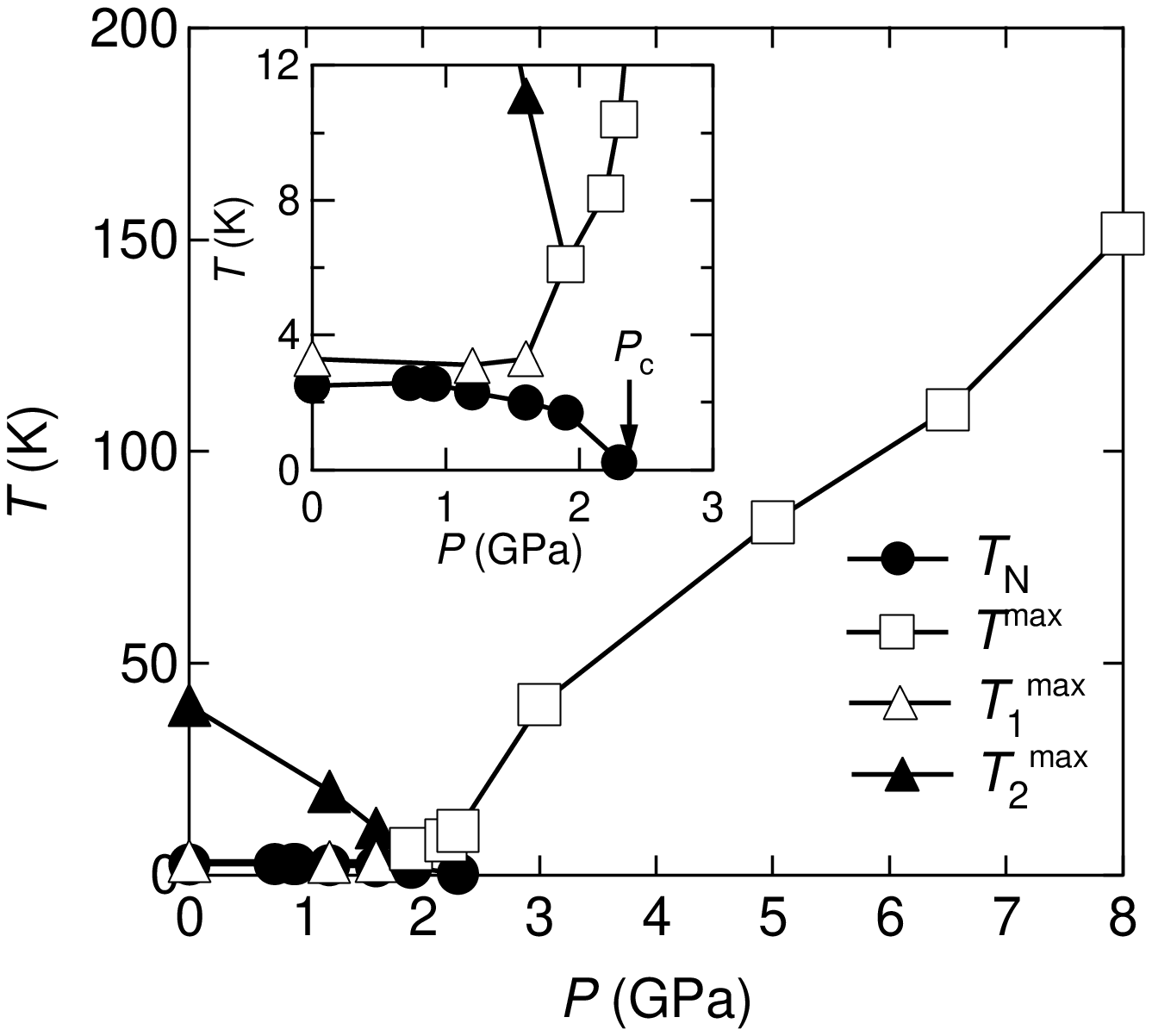}
\caption{
Pressure dependence of the N$\acute{\rm e}$el temperature $T_{\rm N}$, the characteristic temperature $T_{1}^{\rm max}$, $T_{2}^{\rm max}$ and $T^{\rm max}$ obtained from $C_{\rm mag}$($T$) and $\rho_{\rm mag}$($T$). Inset is plotted in an expanded scale in the ranges of 0 $\le  T \le $ 12 K and 0 $\le  P \le $ 3.0 GPa. The solid lines are guides to the eyes. }
\label{fig:6}
\end{center}
\end{figure}
\section{Conclusion}
We have studied on specific heat and electric resistivity under hydrostatic pressure in CeMg$_{2}$Cu$_{9}$ which has the two-dimensional Ce atoms alignment. Magnetic specific heat shows the schottky-type anomaly due to 4$f$ levels of Ce$^{\rm 3+}$ split into three Kramers doublets by 
crystal field effect with energy gap $\Delta_{1}/k_{\rm B}$ = 58 K  and $\Delta_{2}/k_{\rm B}$ = 138 K.   
This crystal field splitting also affect on electric resistivity as the appearance of the two-peaks structure at $T_{1}^{\rm max}$ = 3 K and $T_{2}^{\rm max}$ = 40 K and $-\ln T$ dependencies above $T_{1}^{\rm max}$ and $T_{2}^{\rm max}$.  
Analysis by the Cornut and Coqblin model, we show that the $-\ln T$ dependencies above $T_{1}^{\rm max}$ and $T_{2}^{\rm max}$ are originated from the Kondo effect on the ground and first excited state, respectively. 
With increasing pressure, $T_{2}^{\rm max}$ decreases, and merged with $T_{1}^{\rm max}$ around 1.9 GPa. 
This suggests the Kondo temperature $T_{\rm K}$ becomes same order of the energy level for the crystal field splitting by compression. 
$T_{\rm N}$ decreases with increasing pressure after having maximum 2.6 K at 0.89 GPa, and disappear at the quantum critical point $P_{c} =$  2.4 GPa.   
We also presented the Kondo temperature $T_{\rm K}$ ($\sim$ 8 K) of CeMg$_{2}$Cu$_{9}$. 
\section{Acknowledgments}
This work was partially supported by Grant-in-Aid for COE
Research (No. 13CE2002) and a Scientific Research (B) (No. 13440114) from the Ministry of Education, Culture, Sports, Science and Technology of Japan.

\end{document}